\documentclass[aps,prb,twocolumn]{revtex4}
\usepackage{amssymb}
\usepackage{graphicx}
\usepackage{hyperref}

\begin{document}
\title{Singlet ground state in the spin-1/2 dimer compound Sr$_3$Cr$_2$O$_8$}
\author{Yogesh Singh and D. C. Johnston}
\affiliation{Ames Laboratory and Department of Physics and Astronomy, Iowa State University, Ames, IA 50011}
\date{\today}

\begin{abstract}
Magnetic susceptibility $\chi$ and specific heat $C$ versus temperature $T$ measurements on polycrystalline samples of Sr$_3$Cr$_2$O$_8$ and the isostructural, nonmagnetic compound Sr$_3$V$_2$O$_8$ are reported.  A Curie-Wiess fit to the high-$T$ $\chi(T)$ data for Sr$_3$Cr$_2$O$_8$ indicates that the Cr atoms are in the rare Cr$^{5+}$ (Spin $S$~=~1/2) valence state as expected from the composition.  The ground state was found to be a spin singlet with an excitation gap $\Delta$/k$_{\rm B}$~=~61.9(1)~K to the magnetic triplet states, and a weak interdimer coupling of 6(2)~K was inferred.  The $C$ and $\chi$ measurements on Sr$_3$V$_2$O$_8$ reveal a phase transition at 110~K which is evidently a structural transition.
\end{abstract}

\maketitle

Systems with a spin singlet ground state have been extensively studied because of the various quantum phenomena like Bose-Einstein condensation (BEC) of magnons \cite{Nikuni2000,Waki2004,Jaime2004, Radu2005} and the Wigner crystallization of magnons,\cite{Kageyama1999} that are observed when the spin gap is closed continuously with the application of a magnetic field. The phenomenon of magnon-BEC which was first experimentally observed in TlCuCl$_3$ [\onlinecite{Nikuni2000}] has since been observed in other materials like the quasi-one dimensional (1D) material Pb$_2$V$_3$O$_9$,\cite{Waki2004} and the quasi-2D materials BaCuSi$_2$O$_6$ [\onlinecite{Jaime2004}] and Cs$_2$CuCl$_4$.\cite{Radu2005} On the other hand, the Wigner crystallization of magnons was reported to be realized in the orthogonal dimer system SrCu$_2$(BO$_3$)$_2$ with the Shastry-Sutherland lattice.\cite{Kageyama1999} These discoveries have accelerated the search for new spin dimer systems and new quantum phenomena.

\begin{figure}[t]
\includegraphics[width=3.in]{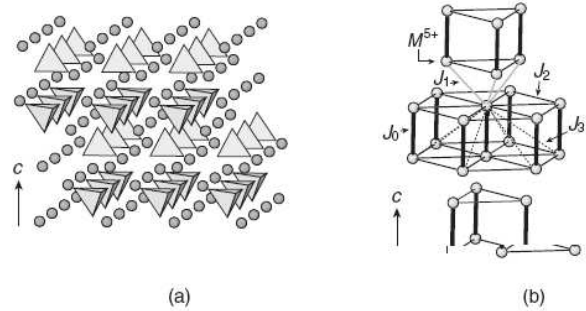}
\caption{(a) The crystal structure\cite{Nakajima2006, Cuno1989} of Sr$_3$Cr$_2$O$_8$ constructed from CrO$_4$ tetrahedra and Sr ions (solid spheres). (b) The arrangement of Cr$^{5+}$ ions (open spheres) in the Sr$_3$Cr$_2$O$_8$ crystal structure.  The shortest Cr-Cr bond is represented by thick black lines.  $J_0$, $J_1$, $J_2$ and $J_3$ denote the first, second, third and fourth nearest-neighbour interactions.  From Ref.\onlinecite{Nakajima2006}    
\label{Figstructure}}
\end{figure}

Recently the compounds Ba$_3$Mn$_2$O$_8$ [\onlinecite{Uchida2001, Uchida2002, Tsujii2005}] and Ba$_3$Cr$_2$O$_8$ [\onlinecite{Nakajima2006}] have been reported to be new spin-1 and spin-1/2 dimer compounds, respectively.  Ba$_3$Mn$_2$O$_8$ exhibits a magnetization plateau at half of its saturation magnetization \cite{Uchida2001, Tsujii2005} like in SrCu$_2$(BO$_3$)$_2$.\cite{Kageyama1999}  The compound Sr$_3$Cr$_2$O$_8$, isostructural with Ba$_3$Cr$_2$O$_8$, was first found in a study of the SrO-Cr oxide phase diagram.\cite{Negas1969} The structure\cite{Cuno1989} is shown in Fig.~\ref{Figstructure}.  However, its magnetic properties have not been investigated. 
We have prepared polycrystalline samples of Sr$_3$Cr$_2$O$_8$ which contains Cr in the unusual 5+ valence state as expected from the composition, and of the isostructural, nonmagnetic compound Sr$_3$V$_2$O$_8$ for comparison.  Herein we report the results of magnetic susceptibility $\chi$ and heat capacity $C$ versus temperature $T$ measurements on these samples.  Analysis of the data indicates that Sr$_3$Cr$_2$O$_8$ is a new spin-1/2 dimer compound with antiferromagnetic exchange couplings, and with a zero-temperature spin gap $\Delta$/k$_{\rm B}$~=~62~K and a small inter-dimer coupling $\approx$ 6~K\@.  Sr$_3$V$_2$O$_8$ undergoes a phase transition below 110~K which is evidently a structural transition.  

Polycrystalline samples of Sr$_3$Cr$_2$O$_8$ and Sr$_3$V$_2$O$_8$ were prepared by solid state synthesis. The starting materials SrCO$_3$ (99.99\%, Alpha Aesar) and Cr$_2$O$_3$ (99.99\%, Alfa Aesar) or V$_2$O$_5$ (99.995\%, MV Labs) were taken in stoichiometric proportion and mixed thoroughly in an agate mortar. The mixture was pressed into a 1/4-inch pellet, placed in a covered Al$_2$O$_3$ crucible and heated in air at 1200~$^\circ$C for 24~hrs and then air quenched to room temperature.  After the initial heat treatment the material was reground and pressed into a pellet and given two similar heat treatments of 24~hrs each with an intermediate grinding after the first 24~hrs.  Hard well-sintered pellets were obtained.  Part of the pellet was crushed for powder x-ray diffraction (XRD).  The XRD patterns were obtained at room temperature using a Rigaku Geigerflex diffractometer with Cu K$\alpha$ radiation, in the 2$\theta$ range from 10 to 90$^\circ$ with a 0.02$^\circ$ step size. Intensity data were accumulated for 5~s per step.  The $\chi(T)$ was measured using a commercial Superconducting Quantum Interference Device (SQUID) magnetometer (MPMS5, Quantum Design) and the $C(T)$ was measured using a commercial Physical Property Measurement System (PPMS5, Quantum Design).  


\begin{figure}[t]
\includegraphics[width=3.in]{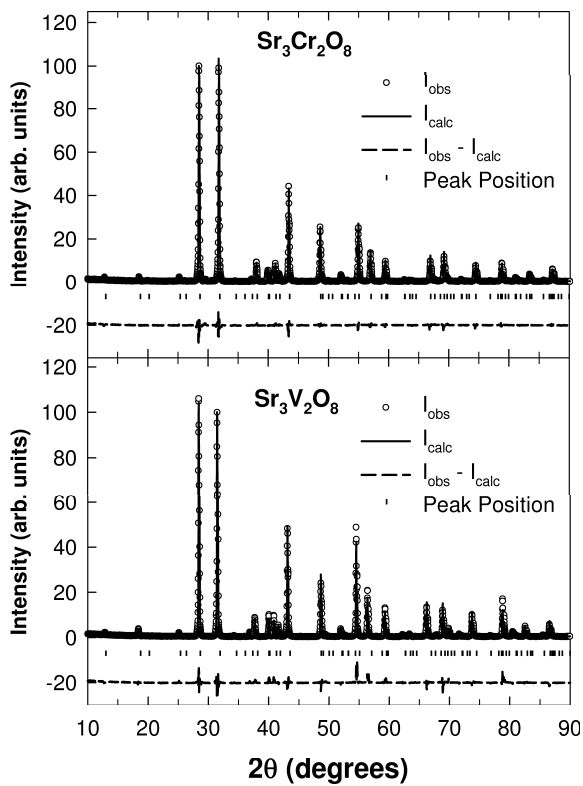}
\caption{Rietveld refinements of the Sr$_3$Cr$_2$O$_8$ and Sr$_3$Cr$_2$O$_8$ X-ray diffraction data. The open symbols represent the observed data, the solid lines represent the fitted pattern, the dotted lines represent the difference between the observed and calculated intensities and the vertical bars represent the peak positions.  
\label{Figxrd}}
\end{figure}

\begin{table*}

\caption{\label{tabStruct}
Structure parameters for Sr$_3$Cr$_2$O$_8$ and Sr$_3$V$_2$O$_8$ refined from powder XRD data.  The overall isotropic thermal parameter $B$ is defined within the temperature
factor of the intensity as $e^{-2B \sin^2 \theta/ \lambda^2}$.}
\begin{ruledtabular}
\begin{tabular}{l|c||cccccc}
Sample & atom & \emph{x} & \emph{y} & \emph{z} & $B$ &$R_{\rm wp}$&$R_{\rm p}$\\
  &  & & && (\AA$^2$) & \\\hline  
Sr$_3$Cr$_2$O$_8$ & Sr~~ & 0 & 0 & 0 & 0.029(2)& 0.152&0.107\\
& Sr~~ &0 & 0 & 0.2034(1) &0.019(1) & & \\
& Cr~~ &0 & 0 & 0.4047(2) & 0.022(2)& & \\
& O~~ &0 & 0 & 0.3233(8) & 0.047(4)&  &\\  
& O~~ &0.8331(7) & 0.1669(7) & 0.8980(3) & 0.039(2)& & \\\hline  
Sr$_3$V$_2$O$_8$ & Sr~~ &0 & 0 &	0 &	0.016(1) & 0.194 &0.147\\
& Sr~~ &0 & 0 & 0.2026(1) & 0.006(1) &  &\\
& V~~ &0 & 0 & 0.4050(3) & 0.009(2) & &\\
& O~~ &0 & 0 & 0.3340(7) & 0.006(4) & &\\  
&O~~ & 0.8400(8)& 0.1600(8) & 0.8955(4) &	0.028(3) && \\
\end{tabular}
\end{ruledtabular}
\end{table*}

All the lines in the X-ray patterns of Sr$_3$Cr$_2$O$_8$ and Sr$_3$V$_2$O$_8$ could be indexed to the known\cite{Cuno1989} trigonal \emph{R$\bar{3}$m} (No.~166) structure and Rietveld refinements,\cite{Rietveld} shown in Fig.~\ref{Figxrd}, of the X-ray patterns gave the corresponding hexagonal lattice parameters $a$~=~$b$~=~5.5718(3) \AA\ and $c$~=~20.1723(13) \AA\ for Sr$_3$Cr$_2$O$_8$, and $a$~=~$b$~=~5.6215(4) \AA\ and $c$~=~20.1152(15)~\AA\ for Sr$_3$V$_2$O$_8$.  These values are in reasonable agreement with previously reported values for single crystalline Sr$_3$Cr$_2$O$_8$ ($a$~=~$b$~=~5.562 \AA, and $c$~=~20.221 \AA )\cite{Cuno1989} and polycrystalline Sr$_3$V$_2$O$_8$ ($a$~=~$b$~=~5.621 \AA, and $c$~=~20.14 \AA).\cite{Durif1959}  Some parameters obtained from the Rietveld refinements of our samples are given in Table~\ref{tabStruct}.

\begin{figure}[t]
\includegraphics[width=3in]{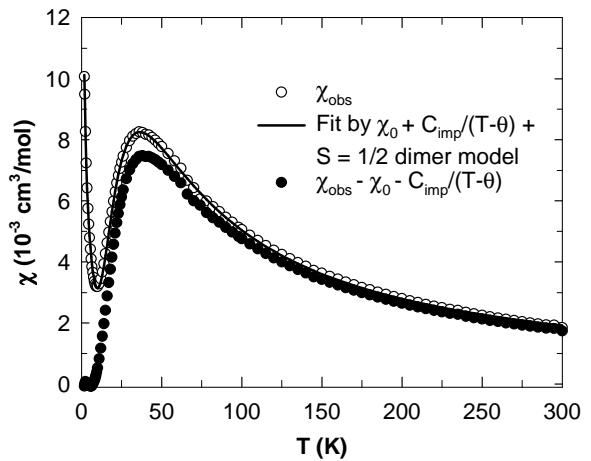}
\caption{The magnetic susceptibility $\chi$ versus temperature $T$ for Sr$_3$Cr$_2$O$_8$\@.  The open circles are the observed susceptibility $\chi_{\rm obs}$, the solid curve is a fit by an isolated dimer model Eq.~(\ref{Eqspindimersus}) and the filled circles are the dimer spin susceptibility obtained by subtracting the constant term and the impurity term in Eq.~(\ref{Eqspindimersus}) from $\chi_{\rm obs}$. 
\label{Figsus}}
\end{figure}

The observed magnetic susceptibility $\chi_{\rm obs}(T)$ for Sr$_3$Cr$_2$O$_8$ is shown in Fig.~\ref{Figsus}.  The $\chi_{\rm obs}(T)$ follows a Curie-Weiss behavior between 100~K and 300~K.  A fit to the data (not shown in the figure) between 100~K and 300~K by the Curie-Weiss expression $\chi = \chi_0+C/(T-\theta$) gives $\chi_0$~=~$-3.6(2)$$\times 10^{-4}$~cm$^3$/mol, $C$~=~0.76(1)~cm$^3$~K/mol and $\theta$~=~$-40(2)$~K\@.  The value of $C$ corresponds to an effective moment of 1.74(1)~$\mu_{\rm B}$/Cr which is close to the value 1.72~$\mu_{\rm B}$ expected for Cr$^{5+}$ ($S$~=~1/2) moments with a $g$-factor of 1.98 taken from a previous EPR measurement.\cite{Gaft2000}  A negative $\theta$ indicates antiferromagnetic interactions between the Cr$^{5+}$ moments.  At lower temperatures $\chi$ exhibits a broad maximum at 38~K before decreasing sharply on cooling.  The $\chi_{\rm obs}(T)$ increases again below 6.5~K and this increase is most likely due to a small amount of paramagnetic impurities.  The strong decrease towards zero in $\chi_{\rm obs}(T)$ below 38~K suggests the presence of a spin-gap.
\noindent
The $\chi_{\rm obs}(T)$ data in the complete temperature range could be fitted by the expression
\begin{eqnarray}
\chi_{\rm obs} &=& \chi_0+\chi_{\rm imp}+\chi_{\rm dimer} \nonumber\\
&=& \chi_0 + C_{\rm imp}/(T-\theta) + {3C/T\over (3+e^{\Delta/k_{\rm B}T})}~~,
\label{Eqspindimersus}	
\end{eqnarray}
\noindent
where $\chi_0$ is a temperature independent term, $\chi_{\rm imp}$ is the contribution from paramagnetic impurities with $C_{\rm imp}$ the Curie constant of the impurities and $\theta$ the Weiss temperature of the impurities, and $\chi_{\rm dimer}$ is the contribution from isolated \emph{S}~=~1/2 dimers.  The dimers have a spin-gap $\Delta$~=~$J_0$ from the singlet ground state to the spin-1 triplet excited states, and a paramagnetic Curie constant $C$.  The fit gave the values $\chi_0$~=~$7(1)\times 10^{-5}$~cm$^3$/mol, $C_{\rm imp}$~=~0.0297(3)~cm$^3$~K/mol, $\theta$~=~$-$1.13(3)~K, $C$~=~0.763(4)~cm$^3$~K/mol and $\Delta$/k$_{\rm B}$~=~61.9(1)~K\@.  The value of $C$ is the same as that found from the above Curie-Weiss fit to the high temperature data.  A very good fit is obtained over the whole temperature range as shown by the solid curve through the $\chi_{\rm obs}$ data in Fig.~\ref{Figsus}.  Also shown in Fig.~\ref{Figsus} are the data after subtracting $\chi_0$ and $\chi_{\rm imp}$ from $\chi_{\rm obs}$, illustrating the spin susceptibility of the dimers.  

To get an estimate of the interdimer interaction strengths $J_1$, $J_2$ and $J_3$ in Sr$_3$Cr$_2$O$_8$ we can write the susceptibility by including these interactions as effective fields.  The contribution to the susceptibility from interacting dimers can then be written as \cite{Johnston1997} 
\begin{equation}
\chi_{\rm interacting Dimer} = {\chi_{\rm Dimer}\over 1+\gamma\chi_{\rm Dimer}} = {3C/T\over (3+e^{\Delta/k_{\rm B}T}+J'/T)}~~,
\label{Eqinteractingdimersus}	
\end{equation}
\noindent
where $\gamma$~=~$J'/3C$ is the molecular field constant, with $J'$~=~$3J_1+6J_2+6J_3$, where $J'$ is the sum of the interdimer exchange interactions.  Fitting $\chi_{\rm obs}$ by the expression
\begin{equation}
\chi_{\rm obs} = \chi_0+\chi_{\rm imp}+\chi_{\rm interacting Dimer}
\label{Eqinteractingdimersus2}	
\end{equation}
\noindent
gives the value $J_0$/k$_{\rm B}$~=~62.0(1)~K for the intradimer exchange interaction, which is very similar to the above value 61.9(1)~K obtained by fitting the data by Eq.~(\ref{Eqspindimersus}), and $J'$/k$_{\rm B}$~=~6(2)~K for the sum of the interdimer exchange interactions.  The rms deviation of the fit from the data is $4.63 \times 10^{-5}\ {\rm cm^3/mol}$ for the isolated dimer model and the slightly smaller value of $4.55 \times 10^{-5}\ {\rm cm^3/mol}$ for the interacting dimer model.  Thus there is little improvement in the fit by including an interaction between dimers.  

The values of $J_0$ and $J'$ indicate that the coupling between the dimers is weak or negligible compared to the coupling within a dimer.  This is in contrast to the case of Ba$_3$Mn$_2$O$_8$ where $J_0$/k$_{\rm B}$~=~17.4~K is comparable to $J'$/k$_{\rm B}$~=~8.3~K [\onlinecite{Uchida2002}] and the case of Ba$_3$Cr$_2$O$_8$ where $J_0$/k$_{\rm B}$~=~25~K is also comparable to $J'$/k$_{\rm B}$~=~7.7~K\@.\cite{Nakajima2006}  The nearest Cr-Cr distance in Sr$_3$Cr$_2$O$_8$ is 3.842 \AA\ whereas the corresponding distance between nearest Cr-Cr in Ba$_3$Cr$_2$O$_8$ is 3.934 \AA\ and between nearest Mn-Mn in Ba$_3$Mn$_2$O$_8$ is 3.984 \AA\@.  The $J_0$ values for the Cr compounds indicate that the intradimer exchange interaction decreases with increasing Cr-Cr distance, as qualitatively expected. 
\begin{figure}[t]
\includegraphics[width=3in]{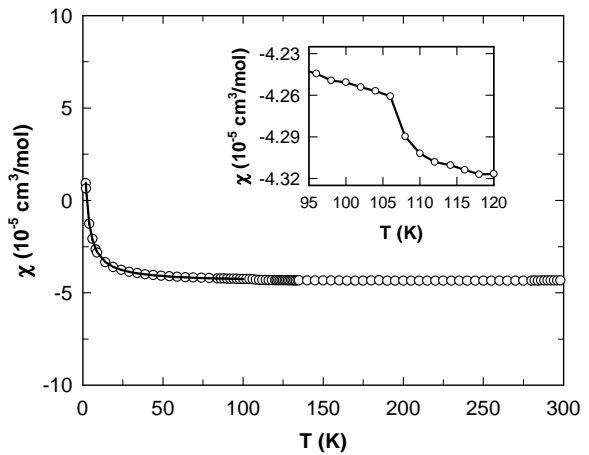}
\caption{The magnetic susceptibility $\chi$ versus temperature $T$ for Sr$_3$V$_2$O$_8$\@.  The solid curve through the data between 1.8~K and 100~K is a fit by the Curie-Wiess law.  The inset shows the data between 95~K and 120~K on an expanded scale to show the transition at 110~K\@.  The solid curve through the data in the inset is a guide to the eye. 
\label{Figsus-SVO}}
\end{figure}

The magnetic susceptibility $\chi$ versus temperature $T$ for Sr$_3$V$_2$O$_8$ is shown in Fig.~\ref{Figsus-SVO} for 1.8~K$\leq T\leq$~300~K\@.  The $\chi(T)$ is almost temperature independent in the whole temperature range.  The inset shows the $\chi$ data between 95~K and 120~K on an expanded scale to highlight a sharp step-like upturn in $\chi$ below 110~K.  This is the onset of a bulk phase transition in Sr$_3$V$_2$O$_8$ as seen in the heat capacity measurement discussed next.  Since there are no magnetic ions in the insulating Sr$_3$V$_2$O$_8$ compound, this phase transition is most likely a structural transition.  The small upturn in the main panel at low temperatures could be fitted (solid curve) by a Curie-Wiess expression $\chi = \chi_0 + C/(T-\theta)$ with the values $\chi_0$~=~$-4.5(1)\times10^{-5}$~cm$^3$/mol, $C$~=~1.68(1)$\times$10$^{-4}$~cm$^3$/mol (corresponding to about 0.2~mol\% spin-1/2 impurities) and $\theta$~=~$-$1.33(2)~K\@.
\noindent

\begin{figure}[t]
\includegraphics[width=3in]{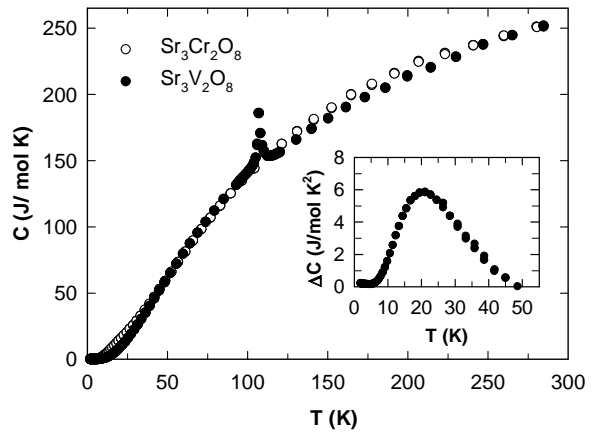}
\caption{The heat capacity $C$ versus temperature $T$ for Sr$_3$Cr$_2$O$_8$ and Sr$_3$V$_2$O$_8$\@.  The inset shows the difference $\Delta C$ between the heat capacities of Sr$_3$Cr$_2$O$_8$ and Sr$_3$V$_2$O$_8$\@.  
\label{FigHC}}
\end{figure}

The heat capacity $C$ versus temperature $T$ of Sr$_3$Cr$_2$O$_8$ and Sr$_3$V$_2$O$_8$ between 1.8~K and 300~K is shown in Fig.~\ref{FigHC}.  There is no signature of any long-range magnetic ordering for Sr$_3$Cr$_2$O$_8$ which rules out the possibility of a magnetic transition at 38~K where the maximum in $\chi(T)$ was observed.  The inset in Fig.~\ref{FigHC} shows $\Delta C$, obtained by subtracting the heat capacity of Sr$_3$V$_2$O$_8$ from that of Sr$_3$Cr$_2$O$_8$.  The $\Delta C$ shows a maximum at about 22~K before decreasing rapidly towards zero at lower temperatures.  The $\Delta C$ becomes negative above 50~K which indicates that the heat capacity for Sr$_3$V$_2$O$_8$ does not quantitatively represent the lattice heat capacity for Sr$_3$Cr$_2$O$_8$.  The reason for this can be seen from Fig.~\ref{FigHC} where a large heat capacity anomaly is seen at about 110~K in the data for Sr$_3$V$_2$O$_8$.  The heat capacity peak occurs at the same temperature at which a step like anomaly was seen in the $\chi$ data for this sample and as stated above most likely arises from a lattice transformation.  This phase transition in Sr$_3$V$_2$O$_8$ at 110~K causes the heat capacity of Sr$_3$V$_2$O$_8$ to cross the data for Sr$_3$Cr$_2$O$_8$ at lower temperatures and leads to the negative $\Delta C$ values above 50~K.  Although the $\Delta C$ may not be a quantitatively correct measure of the magnetic heat capacity of Sr$_3$Cr$_2$O$_8$, the qualitative shape of $\Delta C$ with a peak at about 20~K and a sharp decrease towards zero at low temperatures is what one would expect for the magnetic heat capacity of a spin dimer compound and confirms the presence of a spin gap in this material.  The $C(T)$ data of Sr$_3$V$_2$O$_8$ between 1.8~K and 4~K could be fitted by the expression $C$~=~$\beta T^3$ and gave the value $\beta$~=~0.399(2)~mJ/mol K$^4$.  From the value of $\beta$ one can obtain the Debye temperature $\theta_{\rm D}$ using the expression \cite{Kittel}
\begin{equation}
\Theta_{\rm D}~=~\bigg({12\pi^4Rn \over 5\beta}\bigg)^{1/3}~, 
\label{EqDebyetemp}
\end{equation}
\noindent
where $R$ is the molar gas constant and $n$ is the number of atoms per formula unit (\emph{n}~=~13 for Sr$_3$V$_2$O$_8$).  We obtain $\Theta_{\rm D}$~=~398(2)~K for Sr$_3$V$_2$O$_8$.  The low temperature $C(T)$ of Sr$_3$Cr$_2$O$_8$ has contributions from the paramagnetic impurities which led to an upturn in $\chi(T)$ at low temperatures, and could not be used to obtain $\Theta_{\rm D}$ for this compound.  The upturn in $C(T)$ at low temperatures is most likely due to a splitting of the magnetic ground state of the paramagnetic impurities by the crystalline electric field (CEF).  The CEF splitting would give rise to a Schottky like anomaly in $C(T)$ and the upturn we observe is probably the high temperature tail of such an anomaly.  This maximum of such a Schottky anomaly is expected to shift to higher temperatures on the application of a magnetic field.  We have performed heat capacity measurements at different magnetic fields (not shown here) and we indeed see a broad Schottky-like anomaly peaked at around 5~K in the $C(T)$ data for a field of 1~T\@.  This broad peak shifts upwards in temperature for fields of 5~T and 9~T and merges with the low temperature part of the dimer anomaly.  We were unable to fit the data in all fields to a Schottky model with the same parameters.  This indicates that there might be more than one species of paramagnetic impurities which behave differently.

\noindent

In conclusion, we have synthesized polycrystalline samples of the compounds Sr$_3$Cr$_2$O$_8$ and Sr$_3$V$_2$O$_8$ and studied their structural, magnetic and thermal properties.  Sr$_3$Cr$_2$O$_8$ is a compound with chromium in the rare valence state Cr$^{+5}$ ($S$~=~1/2).  The magnetic properties of Sr$_3$Cr$_2$O$_8$ are consistent with it being a new $S$~=~1/2 spin-dimer compound where the nearest neighbour Cr$^{5+}$-Cr$^{5+}$ ions couple antiferromagnetically to form spin-1/2 dimers with a singlet ground state separated from the excited triplet state by an energy gap $\Delta$/k$_{\rm B}$~=~62~K and a relatively weak interdimer coupling of about 6~K\@.  The data for Sr$_3$V$_2$O$_8$ show that this compound undergoes a phase transition at about 110~K which is most likely a structural transition.  The Debye temperature for Sr$_3$V$_2$O$_8$ is determined to be $\theta_{\rm D}$~=~398(2)~K.
\begin{acknowledgments}
Work at the Ames Laboratory was supported by the Department of Energy-Basic Energy Sciences under Contract No.\ DE-AC02-07CH11358.  
\end{acknowledgments}

\end{document}